\documentclass[11pt]{article}
\usepackage{amsmath}
\usepackage{hyperref}
\usepackage[bf]{caption}
\usepackage{epsfig}
\usepackage{float}
\usepackage{graphicx}
\usepackage{subfig}
\usepackage{color}
\usepackage{amssymb}
\def\ga{\mathrel{\raise.3ex\hbox{$>$\kern-.75em\lower1ex\hbox{$\sim$}}}}
\def\la{\mathrel{\raise.3ex\hbox{$<$\kern-.75em\lower1ex\hbox{$\sim$}}}}

\def\I_M{{I_{\scriptscriptstyle M\times M}}}

\begin{document}

\begin{titlepage}

\thispagestyle{empty}

\vskip 1.2cm

\begin{centering}
\vspace{1cm}
{\Large {\bf Anisotropic branes}}\\

\vspace{1.5cm}
 {\bf Souvik Banerjee $^{\dagger}$}, {\bf Samrat Bhowmick $^{\dagger}$},
 {\bf Sudipta Mukherji $^{\dagger}$}\\
 \vspace{.1in}
 Institute of Physics \\
Bhubaneswar -751 005, India \\

\vspace{3mm}

\end{centering}
\vspace{1.5cm}

\vskip 1.2cm
\vskip 1.2cm
\centerline{\bf Abstract}
\noindent
We present a class of anisotropic brane configurations which shows
BKL oscillations near their cosmological singularities. 
Near horizon limits of these solutions represent Kasner
space embedded in AdS background. Dynamical probe branes
in these geometries inherit anisotropies from the  
background. Amusingly, for a probe M5 brane, we find that there exists
a parameter region where three of its world-volume directions expand
while the rest contract.

\vspace{3.5cm}

\begin{flushleft}
 $^{\dagger}~~$ e-mail address: souvik, samrat, mukherji@iopb.res.in \\

\end{flushleft}
\end{titlepage}
\newpage
\setcounter{footnote}{0}
\noindent

\section*{Introduction}

AdS/CFT correspondence proposes an equivalence between the string theory in AdS space and a conformal field
theory on its boundary. One of the remarkable features of this correspondence is the fact that it relates
a strongly coupled theory with a weakly coupled one. Consequently, it provides us with a way  
to tame the non-perturbative region of one by performing  computations on its dual. Due to strong
gravitational fluctuations, physics around cosmological singularities is dominated
by non-perturbative effects and one hopes that the AdS/CFT correspondence would shed some light
into it. Indeed, in recent years, we  have witnessed several important 
investigations where attempts were made to find the signatures of these singularities
in their gauge theory duals. Expectation is that the dual theory evolution might be able to provide a sensible
quantum description of these singularities. Successes have been varied, please see the references 
\cite{Hertog:2005hu} - \cite{Awad:2009bh}.

Inspired by this line of developments, in this letter, we search for D brane solutions in ten dimensional type IIB theory 
where the world volume metric expands anisotropically and show instabilities within their supergravity descriptions.
We find that appropriately tuning  the 
five form field strength, it is possible to construct a D3 brane with four dimensional Kasner like world volume.
Along with a time-like singularity at $r =0$, the metric shows an additional
cosmological singularity at $t =0$. Perturbation around $t =0$ generates an analogue of 
Belinskii-Lifshitz-Khalatnikov (BKL) oscillations. 
The near horizon geometry of this brane reduces to that of a Kasner-universe in AdS space plus a five sphere along
with an appropriate five form field strength. We further probe the geometry
by a dynamical D3 brane whose world-volume inherits anisotropic expansion/contraction along
with a BKL like oscillation. Similar solutions can be constructed even within eleven dimensional
supergravity. As an illustrative example, we discuss the case of M5 brane. The near horizon 
geometry is now a six dimensional Kasner space. A dynamical probe M5 brane in this space-time
again acquires an anisotropic expansion in some directions and contraction along some.
Amusingly, we find that it is possible to tune parameters in such a manner that three directions expand and
the rest contract. Close to the cosmological singularities supergravity descriptions of all
these solutions are expected to break down. We hope that the gauge theory description would shed some light
on the physics near the singularities.

\section*{D3 brane with anisotropic time-dependent world volume}

Besides the static D branes of odd space dimensions, the IIB string theory admits {\it time dependent}
branes. Consider, for example, the case of 
D3 brane. The equations of motion following from the relevant part of standard IIB supergravity action 
\begin{equation}
 \label{IIB}
 S_{IIB} = -\frac{1}{16 \pi G_{10}}\int d^{10}x \sqrt{-g}
 \left(R - \frac{1}{2}\partial^\mu \phi \partial_\mu \phi - \frac{1}{2\times5!} F_5^2 \right)\;.
\end{equation}
has the following forms:
\begin{eqnarray}
&&R^{\mu}_{\nu} = \frac{1}{2} \partial^\mu \phi \partial_{\nu} \phi + \frac{1}{2 \times 5!} (5 F^{\mu \xi_2...\xi_5}
 F_{\nu \xi_2...\xi_5}- \frac{1}{2} \delta^\mu_\nu F_5^2),\nonumber\\
&&\partial_\mu({\sqrt{g}} F^{\mu \xi_2...\xi_5}) =0, \nonumber\\
&&\nabla^2\phi =0.
\label{2b}
\end{eqnarray}
These equations are solved by
\begin{eqnarray}
&&ds^2 = \Big(1 + \frac{l^4}{r^4}\Big)^{-\frac{1}{2}} \Big[ - dt^2 + t^{2\alpha} dx^2
+ t^{2\beta} dy^2 + t^{2\gamma} dz^2\Big] + \Big(1 + \frac{l^4}{r^4}\Big)^{\frac{1}{2}}
\Big[dr^2 + r^2 d\Omega_5^2\Big],\nonumber\\
&&F_{txyzr} = \frac{4 l^4t^{\alpha+\beta+\gamma}r^3}{(l^4 + r^4)^2},\nonumber\\
&&\phi =0,
\label{timev}
\end{eqnarray}
provided 
\begin{equation}
\alpha + \beta + \gamma = 1~~~{\rm and}~\alpha^2 + \beta^2 + \gamma^2 =1.
\label{restric}
\end{equation}
The numbers $\alpha, \beta, \gamma$ can be organized in an increasing order $\alpha < \beta < \gamma$ and they
vary in the range
\begin{equation}
-\frac{1}{3} \le \alpha \le 0, ~~0 \le \beta \le \frac{2}{3}, ~~{\rm and}~\frac{2}{3} \le \gamma \le 1.
\end{equation}
These numbers can also be parametrized as
\begin{equation}
\alpha(u) = \frac{-u}{1 + u + u^2}, ~~\beta(u) = \frac{1+u}{1 + u + u^2}, ~~\gamma(u) = \frac{u + u^2}{1 + u + u^2},
\end{equation}
where the Lifshitz-Khalatnikov parameter $u \ge 1$. Further, values $u <1$ lead to the same range as
\begin{equation}
\alpha (\frac{1}{u}) = \alpha (u), ~~\beta (\frac{1}{u}) = \beta (u), ~~\gamma (\frac{1}{u}) = \gamma (u).
\end{equation}
The five form charge can be calculated by integrating $*F_5$ over the transverse space and it turns out to
be time independent.

In our convention, the extremal D3 brane is represented by $\alpha = \beta = \gamma =0$ and is not
continuously connected to the above solution. Unlike extremal $D3$ brane, this solution breaks all the supersymmetries 
of IIB theory due to its explicit time dependence. The Kretschmann scalar for the metric is given by
\begin{equation}
R_{\mu\nu\rho\sigma}R^{\mu\nu\rho\sigma} = \frac{16 (-\alpha^2 (l^4 + r^4)^6 + \alpha^3 (l^4 + 
r^4)^6 - 5 l^8 r^4 (l^8 + 12 r^8) t^4)}{r^4 (l^4 +r^4)^5 t^4}.
\end{equation}
In writing the above equation, we have used the condition (\ref{restric}). It has a 
time-like singularity at $r = 0$ at any finite time. 
It, further, has a cosmological 
singularity at $t =0$. 

In the large $r$ limit, equations in (\ref{timev}) reduce to a four dimensional
Kasner solution plus a flat six-dimensional part. Within the Bianchi classification of
homogeneous spaces, the Kasner metric corresponds to choosing all three of
the structure constants to be zero. A generic perturbation near the singularity
breaks these constraints generating Belinskii-Lifshitz-Khalatnikov (BKL) oscillations
\cite{lbk}. To briefly illustrate the BKL oscillation, appropriately generalized to
our context, we replace the world volume metric on the brane by type IX homogeneous space.


To this end, let us consider the brane configuration of the form
\begin{eqnarray}
 \label{BKL-g}
ds^2 = \Big(1 + \frac{l^4}{r^4}\Big)^{-\frac{1}{2}} \big[-dt^2 
&+& (a(t)^2 l_i l_j + b(t)^2 m_i m_j + c(t)^2 n_i n_j) dx^i dx^j\Big]\nonumber\\
&+& \Big(1 + \frac{l^4}{r^4}\Big)^{\frac{1}{2}} \Big[dr^2 + r^2 d\Omega_5^2\Big] \;.
\end{eqnarray}
with the anti-symmetric five form field and the scalar 
\begin{eqnarray}
 \label{BKL-F}
 && F_{txyzr} = \frac{4 r^3 l ^4 a(t) b(t) c(t) \sin (x)}{\left(r^4+l^4\right)^2} \;, \\
 && \phi = 0 \;.
\label{bklt}
\end{eqnarray}
Here $l_i, m_i, n_i$
are frame vectors. For IX metric, all the three structure constants are 1 and the simplest choice for 
the frame vectors is
\begin{equation}
l_i = ({\rm sin} x, - {\rm cos} z ~{\rm sin} x, 0), ~m_i = ({\rm cos} x, {\rm sin} z ~{\rm sin} x, 0),~ n_i = (0, {\rm cos} x, 1).
\end{equation}
The coordinates run through values in the ranges
$0 \le x \le \pi$, $0 \le y \le 2\pi$, $0 \le z \le 4\pi$\footnote{ 
In all our discussion, we will closely follow \cite{ll}. \cite{wiki} also has a lucid review 
of BKL oscillations for types VIII and IX spaces.}. The above configuration (\ref{BKL-g} - 
\ref{bklt}) is
a solution provided they satisfy IIB equations of motion (\ref{2b}). This requirement leads to the
following differential equations for $a, b$ and $c$.
\begin{eqnarray}
 &&\frac{(a_tbc)_t}{abc} = \frac{1}{2a^2b^2c^2}\big[(b^2-c^2)^2-a^4\big] \; ,\nonumber \\
 &&\frac{(ab_tc)_t}{abc} = \frac{1}{2a^2b^2c^2}\big[(c^2-a^2)^2-b^4\big] \; ,\nonumber \\
 &&\frac{(abc_t)_t}{abc} = \frac{1}{2a^2b^2c^2}\big[(a^2-b^2)^2-c^4\big] \; ,\nonumber \\
 &&\frac{a_{tt}}{a}+\frac{b_{tt}}{b}+\frac{c_{tt}}{c} = 0 \; ,
\label{bkltt}
\end{eqnarray}
where the subscript indicates  derivative with respect to $t$. \footnote{For type I spaces, in
which Kasner metric belongs, the right hand sides of all the equations in (\ref{bkltt}) 
would have been zero. This is due to the fact that all the structure constants are zero for 
type I spaces.} These are exactly the equations responsible for generating standard
BKL oscillations. Consequently, the brane world-volume
metric will oscillate with negative powers of $t$ oscillating from one direction to another. In the
next paragraph, for the sake of completeness, we give a brief analysis of this oscillation. 

To proceed, first we notice that if all the expressions on the right hand side of (\ref{bkltt})
are small in some region, the system will have a Kasner-like regime with
\begin{equation}
a \sim t^\alpha, b \sim t^\beta, ~ c \sim t^\gamma,
\end{equation}
where $\alpha, \beta, \gamma$ satisfy constraint as in (\ref{restric}). However, now since 
 $\alpha$ is negative, close to $t = 0$, $a^4$ term in the right hand sides of (\ref{bkltt}) 
will start dominating. It is useful to write these equation in terms of new variables defined
as
\begin{equation}
a = e^p, ~b = e^q, ~c = e^s, ~ e^{p+q+s} d\tau = dt.
\end{equation}
In the vicinity of $t = 0$, (\ref{bkltt}) reduces to
\begin{equation}
p_{\tau \tau} = - \frac{1}{4} e^{4p}, q_{\tau\tau} = s_{\tau\tau} =  \frac{1}{2}e^{4p},
\label{rool}
\end{equation}
where the subscripts $\tau$ indicates derivative with respect to $\tau$. The solution of these equations
should describe the evolution of world-volume metric from the initial state of Kasner metric. In terms
of the new variables, this is equivalent to
\begin{equation}
p_\tau = \alpha, ~q_\tau = \beta, ~s_\tau = \gamma.
\end{equation}
Note that the first equation in (\ref{rool}) can be interpreted as a particle moving 
in the presence of an exponential wall-like potential. Due to the reflection from this barrier,
particle will move with $p_\tau = - \alpha$. However we see from (\ref{rool}) that
$p_\tau + q_\tau$ and $p_\tau + s_\tau$ are constants. So we get
\begin{equation}
q_\tau = \beta + 2 \alpha, ~s_\tau = \gamma + 2 \alpha.
\end{equation}
These lead to
\begin{equation}
e^p = e^{-\alpha \tau}, ~ e^q = e^{(\beta + 2 \alpha)\tau}, ~e^{\tau} = e^{(\gamma + 2 \alpha)\tau}
~{\rm and} ~~ t \sim e^{(1 + 2\alpha)\tau}.
\end{equation}
In terms of the original variables, we can re-write the above as
\begin{equation}
a= t^{\frac{|\alpha|}{1 - 2 |\alpha|}}, ~ b = t^{\frac{\beta-2|\alpha|}{1 - 2|\alpha|}}, ~c = t^{\frac{\gamma - 2|\alpha|}
{1 - 2 |\alpha|}}.
\end{equation}
Therefore the action of the perturbation results oscillations between one Kasner regime to another with
negative power shifting from $a$ to $b$ to $c$, inducing BKL oscillations on the world-volume
of the D3 brane. We should however note that near the curvature singularity, string action  receives
higher derivative gravitational corrections. Consequently, the nature of the singularity and behaviour of its
perturbation may get substantially modified. Solutions may also get modified when one introduces other
matter fields into the theory. If they are represented by a perfect fluids on the world-volume, with
pressure and energy density 
related as $p = \omega \rho$, it can be argued that for $\omega < 1$, BKL oscillations still persist.
However, situation changes drastically for $\omega = 1$, namely for the stiff-matter (a massless scalar
field for example). A general discussion on these issues can be found in \cite{bk, Khalatnikov:2008zt}. 
Indeed it is easy to check that in our previous solution, one can introduce a dilaton with a profile
\begin{equation}
\phi= \lambda \, \log t.
\end{equation}
The metric and the form field remains same as before. However, the exponents now
satisfy new constraint relations : $\alpha + \beta + \gamma = 1, ~~ \alpha^2 + \beta^2 + \gamma^2 = 1 -\lambda^2$. The changes in
these relations allow BKL oscillation for a finite time and the system finally reaches an
attractor.

We now proceed to study the metric in the near horizon limit $r \rightarrow 0$. In this limit, the metric 
reduces to
\begin{equation}
ds^2 = -\frac{r^2}{l^2} dt^2 + \frac{l^2}{r^2} dr^2 + r^2( 
t^{2\alpha} dx^2 + t^{2\beta} dy^2 + t^{2\gamma} dz^2) + l^2 d\Omega_5^2,
\label{kasads}
\end{equation}
with 
\begin{equation}
F_{txyzr} = \frac{4 t r^3}{l^4}, ~~{\rm giving ~potential}~C_{txyz} = \frac{ t r^4}{l^4}.
\end{equation}
The Kretschmann scalar is,
\begin{equation}
R_{\mu\nu\rho\sigma}R^{\mu\nu\rho\sigma} = \frac{16 \alpha^2 (\alpha -1)l^4}{r^4 t^4} - \frac{80}{l^4}.
\end{equation}
We call it a Kasner-AdS space. This Kasner-AdS solution separately satisfies five dimensional Einstein
equation in the presence of a negative cosmological constant and was found in \cite{Frolov:2001wz}
in the context of brane-cosmology.

\section*{Probing with a D3 brane}

In this section, we will probe the geometry (\ref{timev}) with a D3 brane. Distance of the probe
brane from the source now behaves as a scalar whose explicit time dependence can be determined
via a dynamical equation. We take the world-volume directions of the D3 brane as 
$\xi = (t, x, y, z)$. The world-volume action of the D3 brane in background geometry, (\ref{timev}) takes
the form :
\begin{equation}
S=T \int d^4\xi \sqrt{-det \ G_{\alpha\beta}} + T \int d^4\xi \,  \hat{C_4},
\end{equation}
Here $G_{\alpha \beta}$ is the induced metric on the world-volume and $\hat{C_4}$ is the pull-back of the 
background $4$-form potential. $T$ is the brane tension. We turn off all other fields on the brane.\\
The Lagrangian can be cast in the form :
\begin{equation}
L=\sqrt{A(t,r) \ - \ B(t,r) \ \dot{r}^2 \ } - C(r),
\end{equation}
where,
\begin{eqnarray}
\label{ABC}
A(t,r) &=& t^2 \ \left(1+\frac{l^4}{r^4}\right)^{-2} \nonumber \\
B(t,r) &=& t^2 \ \left(1+\frac{l^4}{r^4}\right)^{-1}\nonumber \\
C(r) &=& -\frac{t \ {l^4}}{{l^4} + {r^4}} \nonumber \\
\end{eqnarray}
The equation of motion for $r(t)$ is the Euler-Lagrange equation derived from (\ref{ABC}) :
\begin{eqnarray}
\label{eomr}
 && r^4\big(l^4+r^4\big)^2[t\ddot{r}+\dot{r}]  -(l^{4}+r^{4})^3 \dot{r}^3
 -2 t l^4 r^3\left[ 3(l^4+r^4)\dot{r}^2 - 2 r^4\right]
 \nonumber \\
 &&  -4 t l^4 r \left[r^4 - \left(l^4+r^4\right)\dot{r}^2  \right]^{\frac{3}{2}} = 0
\end{eqnarray}
Here dot represents derivative with respect to $t$.
Once this equation is solved with appropriate boundary conditions,
the metric on the probe brane is uniquely determined. We will carry out this computation in this section.
However, owing to the explicit time dependence in the background geometry, we find that the dynamical
equation can not be solved analytically. Fortunately, it is not hard to find numerical solution and
a typical behaviour is shown in the figure 1.

\begin{figure}[H]
\label{r-tplot}
\centering
\includegraphics[width=.75\textwidth]{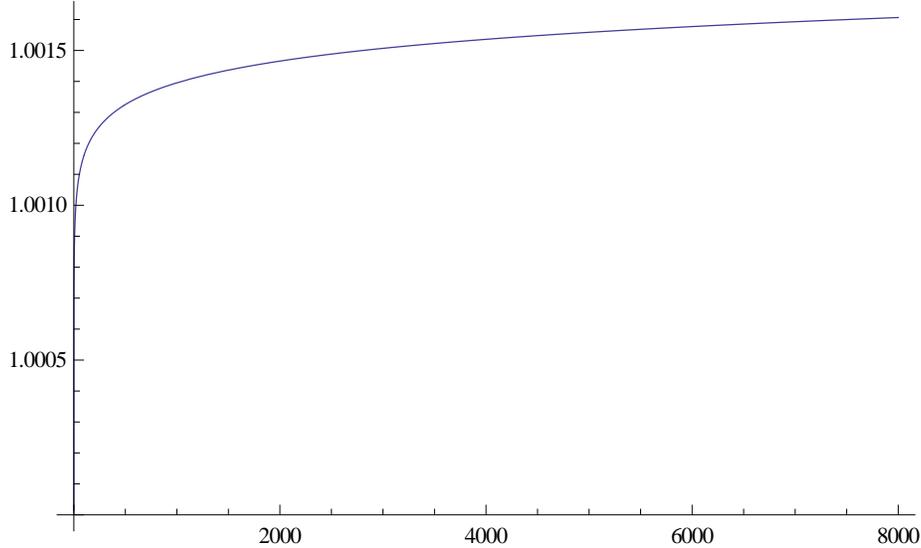}
\caption{Plot of $r$ as a function of time, $t$.}
\end{figure}

The functions that govern the anisotropic expansions in three spatial directions are $t^\alpha \ f(r)$, 
$t^\beta \ f(r)$, $t^\gamma \ f(r)$, where 
\begin{equation}
\label{scale-factor}
f(r) = \left(1+\frac{l^4}{r^4}\right)^{-\frac{1}{4}}.
\end{equation}
In order that the near horizon geometry is an $AdS$, as mentioned earlier, $\alpha$,$\beta$,$\gamma$ must 
satisfy the constraint :
$\alpha^2 + \beta^2 + \gamma^2 = \alpha + \beta + \gamma = 1$.
This means, once we specify one of the three,
say, $ \alpha $, the other two are automatically fixed : 
\begin{eqnarray}
\label{alpha}
\beta  &=& \frac{1}{2} \left(1 - \alpha +\sqrt{-3 \alpha ^2+2 \alpha +1}\right) \nonumber \\
\gamma &=& \frac{1}{2} \left(1 - \alpha -\sqrt{-3 \alpha ^2+2 \alpha +1}\right)
\end{eqnarray}

Ideally, in cosmology, one defines cosmological time, $\eta$ with which the metric on the probe brane takes 
the form :
\begin{equation}
dS_{brane}^2 = -d\eta^2 + \left(1+\frac{l^4}{r^4 (\eta)}\right)^{-\frac{1}{2}} (t^{2 \alpha}(\eta) dx^2 + 
t^{2 \beta}(\eta) dy^2 + t^{2 \gamma}(\eta) dz^2).
\end{equation}
with
\begin{equation}
\frac{d \eta}{d t} =\sqrt{\left(1+\frac{l^4}{r^4}\right)^{-\frac{1}{2}} - 
\left(1+\frac{l^4}{r^4}\right)^{\frac{1}{2}}\left(\frac{d r(t)}{dt}\right)^2}.
\end{equation}
The behaviour of time, $t$ as a function of $\eta$ is depicted in figure 2.

\begin{figure}[H]
\label{t-eta}
\centering
\includegraphics[width=.5\textwidth]{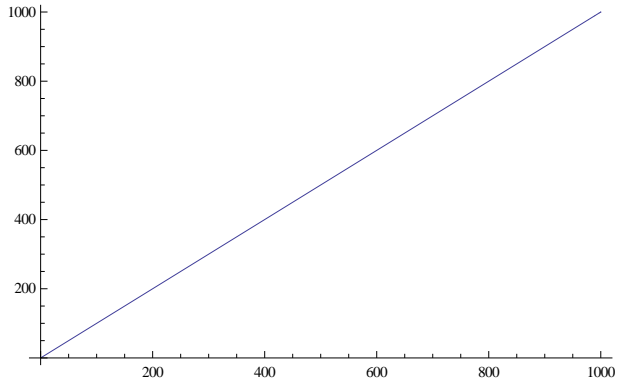}
\caption{ 
Plot of $t$ as a function of $\eta$.}
\end{figure}
At this, we plot the functions  $t^\alpha \ f(r)$, $t^\beta \ f(r)$, $t^\gamma \ f(r)$ as functions 
of $\eta$ parametrically. Here $f(r)$ is defined through (\ref{scale-factor}). 
One can tune the values of $\alpha$, $\beta$, $\gamma$ consistent with 
the Kasner constraints so that one of them goes down to zero (decelerating) while two of them go up 
(accelerating) with cosmic time and vice versa. One such plot is given in figure 3.

\begin{figure}[H]
\label{scalefactor-eta}
\centering
\includegraphics[width=.62\textwidth]{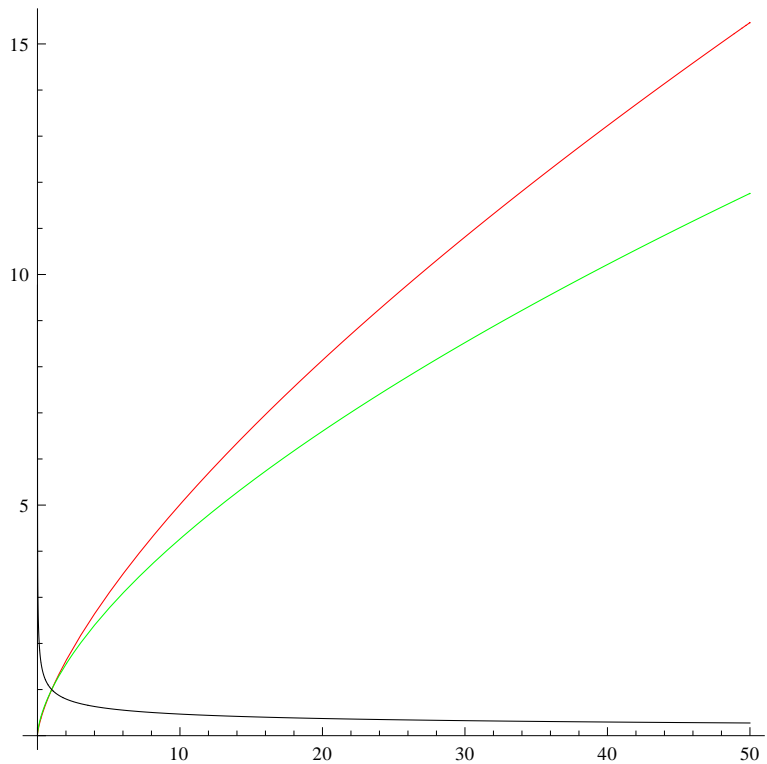}
\caption{ The functions, $f_1=t^\alpha(\eta) \ f(r)$, $f_2=t^\beta(\eta) \ f(r)$, 
$f_3=t^\gamma(\eta) \ f(r)$, with $f(r)$ given in (\ref{scale-factor})
are plotted as functions of $\eta$. 
$\alpha$, $\beta$ and $\gamma$ 
are 0.7, .632, -0.332 respectively
The plot of $f_1$ is in red, and that of $f_2$ and $f_3$
are in green and black respectively.}
\end{figure}

\section*{The  M5 Brane}
Our previous discussion can easily be extended to eleven dimensions. Here we discuss the case of a M5 brane.
We start with $d=11$ supergravity action 
\begin{equation}
 \label{11dsugra}
 S_{11d} = -\frac{1}{2 \ \kappa_{11}^2}\int d^{11}x \sqrt{-g}
 \left(R - \frac{1}{48}  \ F_4^2\right)\;,
\end{equation}
which is a generic action for the bosonic part of $d=11$ supergravity so long as we concentrate on static, flat 
translationally invariant p-brane solutions.

The equations of motion arising from 
(\ref{11dsugra}) admits a solution of the form : 

\begin{eqnarray}
\label{metric11d}
ds^2 &=& \left(1 + \frac{l^3}{r^3}\right)^{-\frac{1}{3}} \left[-dt^2 + t^{2 \alpha_1^2} dx_1^2 
+ t^{2 \alpha_2^2} dx_2^2 + t^{2 \alpha_3^2} dx_3^2 
+ t^{2 \alpha_4^2} dx_4^2 + t^{2 \alpha_5^2} dx_5^2\right] \nonumber \\
&+& \left(1 + \frac{l^3}{r^3}\right)^{\frac{2}{3}} \left[dr^2 + r^2 d\Omega_4^2\right],
\end{eqnarray}
along with
\begin{equation}
\label{gauge11d}
F_{t x_1 x_2 x_3 x_4 x_5 r} = \frac{3 \ l^3 \ t \ r^2}{(l^3 + r^3)^2}
\end{equation}

provided $\alpha_1 + \alpha_2 + \alpha_3 + \alpha_4+ \alpha_5 = 1$ and 
$\alpha_1^2 + \alpha_2^2 + \alpha_3^2 + \alpha_4^2+ \alpha_5^2 = 1$.

In the near horizon limit, i.e. $r\rightarrow 0$, the metric and the non-zero component of the 
form field reduce to the forms :
\begin{eqnarray}
\label{nehor} 
ds^2 &=& \frac{r}{l} \left[-dt^2 + t^{2 \alpha_1^2} dx_1^2 
+ t^{2 \alpha_2^2} dx_2^2 + t^{2 \alpha_3^2} dx_3^2 
+ t^{2 \alpha_4^2} dx_4^2 + t^{2 \alpha_5^2} dx_5^2\right] \nonumber \\
&+& \frac{l^2}{r^2} \left[dr^2 + r^2 d\Omega_4^2\right],\nonumber  \\
F_{t x_1 x_2 x_3 x_4 x_5 r} &=& \frac{3 \ t \ r^2}{l^3},
\end{eqnarray}
and hence the potential is given by $C_{t x_1 x_2 x_3 x_4 x_5} = \frac{t \ r^3}{l^3}$. 

We now make the following change of coordinates :
\begin{equation}
w^2 = \frac{r}{l^3}. 
\end{equation}
With this, the metric in (\ref{nehor}) takes the form :
\begin{equation}
\label{nehor1} 
ds^2 = \frac{w^2}{4 l^2}\left(-d\bar{t}^2 + \bar{t}^{2 \alpha_1^2} d\bar{x}_1^2 
+ \bar{t}^{2 \alpha_2^2} d\bar{x}_2^2 + \bar{t}^{2 \alpha_3^2} d\bar{x}_3^2 
+ \bar{t}^{2 \alpha_4^2} d\bar{x}_4^2 + \bar{t}^{2 \alpha_5^2} d\bar{x}_5^2 \right) + 
4 \ l^2 \frac{dw^2}{w^2} + l^2 d\Omega_4^2,
\end{equation}
where $\bar{x}_i$ and $\bar{t}$ are suitably scaled versions of the coordinates, ${x}_i$ and ${t}$ 
respectively. It is worth mentioning in this regard that the scaling of the coordinates
will not be the same because of the presence of different powers of $t$ in front of $dx_i^2$.
This is a consequence of anisotropy.

Following our nomenclature, (\ref{nehor1}) is a metric of seven dimensional Kasner--Ads space
plus a four sphere. For
$\alpha_i = 0$ for $i = 1,...,5$, this reduces to our known 
$AdS_7(2L) \times S^4(L)$ solution.

\subsection*{Probing with a M5 brane}
In the same spirit as we considered the case of probe D3 brane, we now consider a probe M5 brane in the
background (\ref{metric11d}) and (\ref{gauge11d}). 

In PST formalism \cite{Pasti:1996vs}, the world-volume action of M5 brane is given in terms of a gauge 
invariant 
$3$-form field strength, ${\cal H}^{(3)} = d{\cal A}^{(2)} + {\cal C}^{(3)}$, where ${\cal A}^{(2)}$ is 
world-volume $2$-form and ${\cal C}^{(3)}$, target space $3$-form. The world volume action in this 
formalism is written as :

\begin{equation}
S_{M5} = T_{M5} \int d^6 \xi \left[{\cal L}_{DBI} + {\cal L}_{KE} + {\cal L}_{WZ}\right],
\end{equation}
where 
\begin{eqnarray}
{\cal L}_{DBI} &=& \sqrt{-det\left(G_{ij} + {\tilde{\cal H}}_{ij}\right)} \quad {\text {is the Dirac-Born-
Infeld Lagrangian,}}\nonumber \\
{\cal L}_{KE} &=& \frac{1}{24 \left(\partial a\right)^2} \epsilon^{ijklmn} {\cal H}_{lmn} {\cal H}_{jkp}
G^{pq} \partial_i a \partial_q a \quad {\text {is the kinetic piece for the $3$-form,}} \nonumber \\
{\cal L}_{WZ} &=& \frac{1}{6!} \epsilon^{ijklmn} \left[{\cal C}^{(6)}_{ijklmn} + 10 \ {\cal H}_{ijk} \
{\cal C}^{(3)}_{lmn}\right] {\text{is the Wess-Zumino term.}}\nonumber \\
\end{eqnarray}
Here $S_{\alpha \beta}$ is the induced metric on the world-volume, ${\cal C}^{(3)}$ and ${\cal C}^{(6)}$
are the pull-backs of the $3$-form and $6$-form background potentials respectively. ${\tilde{\cal H}}$ is
defined as 
\begin{equation}
{\tilde{\cal H}}^{ij} = \frac{1}{3! \sqrt{-det G} \ \sqrt{-(\partial a)^2}}  \epsilon^{ijklmn}
\partial_k a \ {\cal H}_{lmn}.
\end{equation}
``$a$'' is an auxiliary scalar field introduced in PST formalism to maintain manifest covariance. \\

If we now take the world-volume directions of the M5 brane as 
$\xi = (t,x_1, x_2, x_3, x_4, x_5)$,
it can be explicitly checked that, in this ``static gauge'', there will be no component of ${\cal C}^{(3)}$
in world-volume directions. We further simplify the system by turning off the world-volume $2$-form, 
${\cal A}^{(2)}$. With all these taken into account, the full Lagrangian takes the simple form :
\begin{equation}
L=\sqrt{A(t,r) \ - \ B(t,r) \ \dot{r}^2 \ } - C(r),
\end{equation}
where,
\begin{eqnarray}
\label{ABC_M5}
A(t,r) &=& t^2 \ \left(1+\frac{l^3}{r^3}\right)^{-2} \nonumber \\
B(t,r) &=& t^2 \ \left(1+\frac{l^3}{r^3}\right)^{-1}\nonumber \\
C(r) &=& -\frac{t \ {l^3}}{{l^3} + {r^3}} \,.\nonumber \\
\end{eqnarray}
Here dot represents derivative with respect to $t$.
The Euler Lagrange equation for $r(t)$ is :
\begin{eqnarray}
\label{eom}
&&2r^4 \left(l^3+r^3\right)^2 \left[t\ddot{r}+\dot{r}\right] -2r \left(l^3+r^3\right)^3 \dot{r}^3 
-3 t l^3r^3 \left[3 \left(l^3+r^3\right) \dot{r}^2-2r^3\right] \nonumber\\
&&-6 t l^3 r^{\frac{3}{2}}\left[r^3 - (l^3+r^3) \dot{r}^2\right]^{\frac{3}{2}} = 0 \;.
\end{eqnarray}

In order to draw a cosmological interpretation of the solutions we obtain from (\ref{eom}),
as usual, we go to the ``cosmic time'' coordinate, $\eta$, in which the metric on the
brane assumes a form :

\begin{equation}
ds_{{\it brane}}^2 = -d{\eta(t)}^2 + \left(1+\frac{l^3}{r^3 (\eta)}\right)^{-\frac{1}{3}}
\left(\sum_{i=1}^5{t}^{2 \alpha_i}(\eta) d{x}_i^2\right) \, ,
\end{equation}
with
\begin{equation}
\frac{d \eta}{d t} =\sqrt{\left(1+\frac{l^3}{r^3}\right)^{-\frac{1}{3}} - 
\left(1+\frac{l^3}{r^3}\right)^{\frac{2}{3}}\left(\frac{d r(t)}{dt}\right)^2}.
\end{equation}
The functions that govern the expansion of the universe in the spatial world-volume directions of the 
brane are in this case $t^{\alpha_i} \ f(r)$, where
\begin{equation}
\label{scalefactor-M5}
 f(r) = \left(1+\frac{l^3}{r^3}\right)^{-\frac{1}{6}} \;.
\end{equation}

We can choose $\alpha_i$'s so that three of them are the same and mimics isotropic expansion in three
directions. The other two are anisotropic. Such a situation can be parametrized as :
\begin{eqnarray}
\alpha_1 &=& \alpha_2 = \alpha_3 = p \nonumber \\
\alpha_4 &=& \frac{1}{2} \left(\sqrt{-15 p^2+6 p+1}-3 p+1\right) \nonumber \\
\alpha_5 &=& \frac{1}{2} \left(-\sqrt{-15 p^2+6 p+1}-3 p+1\right) \, .
\end{eqnarray}
Interestingly there exists a narrow window of parametric value for $p$, in which $\alpha_i$ for $i=1, 2, 3$ 
are positive and $\alpha_4$ and $\alpha_5$ are negative. An illustrative plot is shown in figure 4 for a particular
value of $p$.

\begin{figure}[H]
\label{metaplot}
\centering
\includegraphics[width=.65\textwidth]{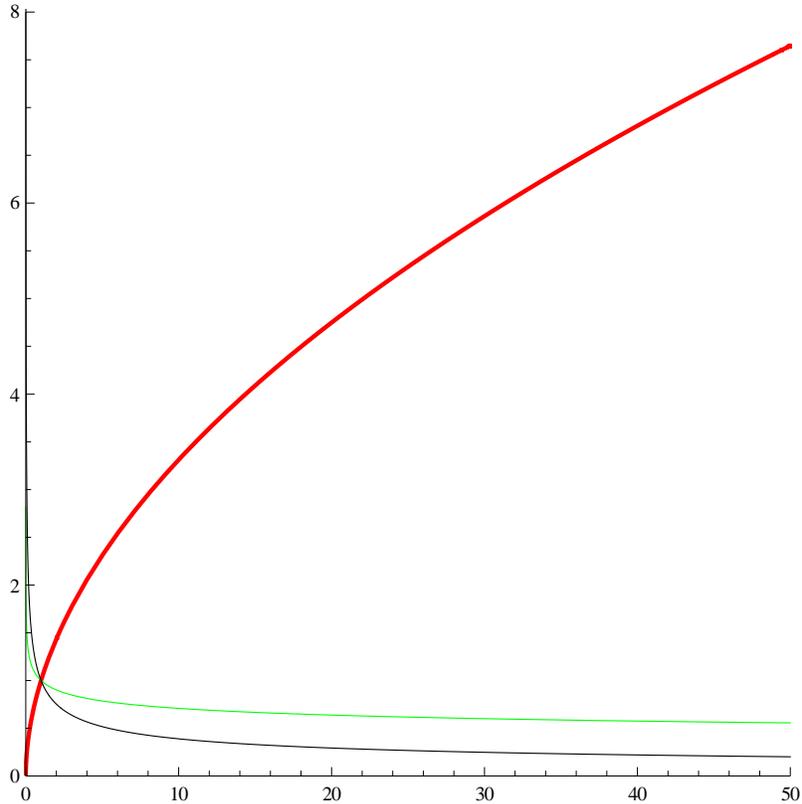}
\caption{ The functions, $t^{\alpha_i}(\eta) \ f(r)$, for $i = 1,2,\cdots,5$
are plotted as functions of $\eta$ for $p=0.52$.
This corresponding values for $\alpha_i$'s are : $\alpha_1 = \alpha_2 = \alpha_3 = 0.52$ , 
$\alpha_4 = -0.15$ and $\alpha_5 = -0.41$. The isotropic expansion corresponding to 
$\alpha_1$, $\alpha_2$, $\alpha_3$ is plotted in red. The contraction corresponding to 
$\alpha_4$, $\alpha_5$ are plotted in green and black respectively.}
\end{figure}



To conclude, we have presented a class of brane configurations which shows BKL oscillations 
near their cosmological singularities. It will be worthwhile to look for the signatures of
these oscillations in their dual descriptions. We hope to report on this in the near future.


\section*{Acknowledgments}

We are grateful to Sumit Das for a discussion at an early stage of this work. We also thank
Sumit Das and K. Narayan for their comments on a previous version of this manuscript.


\end{document}